\newcommand{\CLASSINPUTinnersidemargin}{15mm}
\newcommand{\CLASSINPUToutersidemargin}{14mm}
\newcommand{\CLASSINPUTtoptextmargin}{20mm}
\newcommand{\CLASSINPUTbottomtextmargin}{25mm}
\documentclass[journal,twocolumn]{IEEEtran} %!PN

\usepackage{times}

\usepackage{graphicx}

\usepackage{amsmath}
\usepackage{times}
\usepackage{graphicx}
\usepackage{amsfonts}
\usepackage{amssymb}
\usepackage{cite}
\usepackage{subfigure}
\usepackage{url}
\usepackage{flushend}
\usepackage[nolist,nohyperlinks]{acronym}
\usepackage[numbers,sort&compress]{natbib}

\DeclareGraphicsExtensions{.png,.eps,.ps,.pdf}

\usepackage{tikz}
\usetikzlibrary{arrows,	% Tipos de flechas
				shapes, % Formas de bloques
				positioning, % Posicionamiento de nodos 
				fit,	% Ajustar a nodos
                backgrounds, % Fondos
                mindmap,
                petri,
				intersections, % Intersecciones entre formas o curvas
				external % Externalizar figuras tikz
				}
\definecolor{emerald}{RGB}{69,155,61}
\definecolor{gold}{RGB}{244,216,51}
\definecolor{pink}{RGB}{235,44,206}
\usepackage{verbatim}

\begin{acronym}
\acro{SNR}{signal-to-noise ratio}
\acro{AoF}{amount of fading}
\acro{OP}{outage probability}
\acrodefplural{SNR}[SNRs]{signal-to-noise ratios}
\acro{RV}{random variable}
\acrodefplural{RV}[RVs]{random variables}
\acro{LOS}{line-of-sight}
\acrodefplural{LoS}{Line-of-sight}
\acro{EH}{Energy Harvesting}
\acro{CSI}{channel state information}
\acro{PDF}{probability density function}
\acro{CDF}{cumulative distribution function}
\acrodefplural{CDF}[CDFs]{cumulative distribution functions}
\acro{IoT}{Internet of Things}
\acro{PB}{Power Beacon}
\acrodefplural{PB}[PBs]{Power Beacons}
\acro{WPC}{Wireless Powered Communication}
\acro{WPT}{Wireless Power Transmission}
\acro{MC}{Monte Carlo}
\acro{PLS}{Physical Layer Security}
\acro{AWGN}{Additive White Gaussian Noise}
\acro{SWIPT}{Simultaneous Wireless and Information Power Transfer}
\end{acronym}

\makeatletter
\def\blfootnote{\xdef\@thefnmark{}\@footnotetext}
\makeatother

\newcommand{\MATLAB}{\textsc{Matlab}}
\newcommand{\MATHE}{\textsc{Mathematica}}

\tikzstyle{int}=[draw, fill=cyan!20, minimum size=2em]
\tikzstyle{int_blue}=[draw, fill=blue!20, minimum size=2em]
\tikzstyle{int_green}=[draw, fill=green!20, minimum size=2em]
\tikzstyle{int_red}=[draw, fill=red!20, minimum size=2em]
\tikzstyle{init} = [pin edge={to-,thin,black}]

\renewcommand\IEEEkeywordsname{Index Terms}

\begin{document}

\title{{Effect of Correlation between Information and Energy Links in Secure Wireless Powered Communications}}
%The Role of Correlation between Energy and Information Links in Secure Wireless Powered Communications

\author{Antonio Tarr\'ias-Mu\~noz, Jos\'e Luis Matez-Bandera, Pablo Ram\'irez-Espinosa and F. Javier L\'opez-Mart\'inez}

%\author{
%%\authorblockN{Lista de autores}
%\IEEEauthorblockN{Antonio Tarr\'ias-Munoz$^{*}$, Jos\'e Luis Matez-Bandera$^{+}$, Pablo Ram\'irez-Espinosa$^o$, F. Javier L\'opez-Mart\'inez$^{*}$}
%\authorblockA{antoniotarriasm@uma.es, josematez@uma.es, pres@es.aau.dk, fjlopezm@ic.uma.es}
%\authorblockA{$^{*}$Dpto. de Ingenier\'ia de Comunicaciones, Universidad de M\'alaga, ETSI Telecomunicaci\'on, 29071 M\'alaga.}
%\authorblockA{$^{+}$Dpto. de Ingenier\'ia de Sistemas y Autom\'atica, Universidad de M\'alaga,  ETSI Inform\'atica, 29071 M\'alaga.}
%\authorblockA{$^{o}$Connectivity Section, Department of Electronic Systems, Aalborg University, Aalborg Øst 9220, Denmark}
%}

\maketitle

\blfootnote{\noindent This work has been submitted to the IEEE for possible publication. Copyright may be transferred without notice, after which this version may no longer be accessible. Manuscript received April xx, 2020; revised XXX. This work has been funded by the Spanish Government and the European Fund for Regional Development FEDER (project TEC2017-87913-R) and the ``Becas Colaboraci\'on con departamentos'' program, by Junta de Andalucia (project P18-RT-3175, TETRA5G) by University of M\'alaga. The review of this paper was coordinated by XXXX.}
\blfootnote{\noindent A. Tarr\'ias-Mu\~noz and F. J. L\'opez-Mart\'inez are with Departamento de Ingenieria  de Comunicaciones, Universidad de Malaga - Campus de Excelencia Internacional Andalucia Tech., Malaga 29071, Spain (e-mail: $\rm antoniotarriasm@uma.es$ and $\rm fjlopezm@ic.uma.es$).}

\blfootnote{\noindent J.L. Matez-Bandera is with Dpto. de Ingenier\'ia de Sistemas y Autom\'atica, Universidad de M\'alaga - Campus de Excelencia Internacional Andalucia Tech., Malaga 29071, Spain (e-mail: $\rm josematez@uma.es$).}

\blfootnote{\noindent P. Ram\'irez-Espinosa is with the Connectivity Section,  Department of Electronic  Systems, Aalborg University, Aalborg {\O}st 9220, Denmark (e-mail: $\rm pres@es.aau.dk$).}

\blfootnote{Digital Object Identifier 10.1109/XXX.2019.XXXXXXX}

\begin{abstract}
In this paper, we investigate the impact of correlation between the energy and information links in wireless power transfer systems, from a physical layer security perspective. With that aim, we first determine how correlation can affect system capacity in practical energy harvesting set-ups in the absence of eavesdroppers. We quantify that even though link correlation improves the average signal-to-noise ratio (SNR) for a fixed transmit power, it also increases its variance, which ultimately degrades capacity. Based on this observation, we show that correlation between the energy and information links may be detrimental/beneficial for the secrecy capacity in the high/low legitimate SNR regime, whenever such correlation affects the legitimate user. Conversely, we also point out that when link correlation for the wiretap link is rigorously accounted for, it barely affects secrecy performance, causing only a minor degradation in some instances.
\end{abstract}

\medskip

\begin{IEEEkeywords}
Secrecy capacity, outage probability, eavesdropper, wireless power transfer, capacity, correlation.
\end{IEEEkeywords}

\section{Introduction}
Over the last years, the number of devices connected to wireless networks has been dramatically increasing due to the advent of 5G technology \cite{5G} and the development of the \ac{IoT} \cite{IoT2}. In this line, \ac{EH} technology \cite{EH1,EH2}, which allows devices to harvest the energy required for operation in a wireless fashion, is one of the key enablers for massive wireless sensor deployment. Perhaps the most feasible use case in this scenario is \ac{WPT} \cite{Liu2013, Xia2015}, which employs dedicated \acp{PB} to wirelessly convey energy to the network agents.

\ac{WPC} systems \cite{Bi2016} integrate \ac{WPT} technology with traditional wireless communications, operating in a number of forms which may include \ac{SWIPT} alternatives. In many cases, \ac{WPT}-based solutions imply a bidirectional operation between the \ac{EH} node and the \ac{PB}: for instance, the \ac{PB} wirelessly transmits energy to the \ac{EH} node to provide it with enough power as to report some information back to the \ac{PB}. This is also the conventional mode of operation of backscatter communication systems \cite{Han2017}.  In both instances, this translates into a clear \ac{LOS} between the system agents, which combined with the slow variability of fading in these scenarios implies a non-negligible correlation between the information and energy transfer links \cite{Griffin07,Zhang19}.

The effect of correlation between the energy and information links in \ac{WPC} systems has been scarcely taken care of in the literature, being the only available references focused on backscatter communication systems \cite{He2011,Gao2016,Zhang19,Matez2020}. Such lack of reference works is even intensified in the context of \ac{PLS} \cite{Bloch2008}. In this scenario, communication between legitimate peers is observed by an external eavesdropper, which complicates the problem under analysis because of adding a new wireless link. Indeed, the correlation between the desired and eavesdropper's links has been analyzed in the literature \cite{Jeon11,Sun12}; however, the effect of correlation between \emph{energy} and \emph{information} transfer links in the context of \ac{WPC} has only been partially addressed in \cite{Zhang19b}. In this work, we aim to shed light on this issue, with one key question in mind: \emph{is correlation between the information and energy links beneficial in some way for physical layer security in \ac{WPC}?} This is a rather challenging problem to be solved from a mathematical viewpoint, for which we provide a clear procedure to derive the key secrecy performance metrics for an arbitrary choice of fading distribution, as well as some specific examples for the case of Rician and Rayleigh fading. Besides, the effect of correlation on the average \ac{SNR} at the receiver ends needs to be correctly accounted for; we will later see that whenever the interaction between the average \ac{SNR}s at the legitimate and eavesdropper's sides is neglected, the conclusions about the effect of correlation could be incorrect in some instances.

The remainder of this paper is structured as follows: the system set-up under consideration is introduced in Section \ref{systemModel}. In Section \ref{CorrelationWPC}, we first examine the effect of correlation in a WPC system in the absence of eavesdroppers. After that, in Section \ref{CorrelationCS} we analyze the effect of correlation in \ac{PLS} in different practical scenarios. Finally, in Section \ref{Conclusions} the main conclusions are summarized.

\section{System Model}
\label{systemModel}

\begin{figure}[t]
\centering
\begin{tikzpicture}[node distance=3.5cm,auto,>=latex']
    \node [int, pin={[init]below:$P_0=g\left(P_{I}\right)$}] (a) {$A$};
    \node [int_green, pin={[init]below:$P_{T}$}, left of=a,node distance=3cm] (b) {$\rm{PB}$};
    \node [int] (c) [right of=a] {$B$};
    \node [int_red] (e) [below of=c, node distance=1.5cm] {$E$};
    \node [coordinate] (end) [right of=c, node distance=2cm]{};
    %\path[->] (b) edge node {$a$} (a);
    \path[->] (a) edge node {$h_B$} (c);
    %\draw[->] (c) edge node {$p$} (end) ;
    \path[->] (a) edge node {$h_E$} (e);
    \path[->] (b) edge node {$h_{P}$} (a);
\end{tikzpicture}
\caption{Wireless power transfer-based system set-up}
\label{PB_system}
\end{figure}
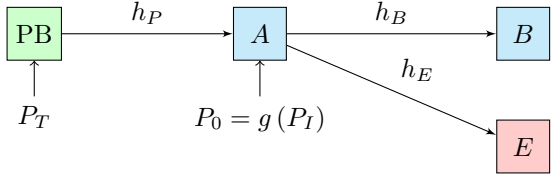

%\subsection{General Set-up}
%\label{generalSetup}
We consider the system set-up in Fig \ref{PB_system}. A legitimate user (Alice) operates thanks to the energy wirelessly conveyed by a remote \ac{PB}. Now, Alice wants to communicate with a legitimate user (Bob) over a wireless fading channel, in the presence of a non-legitimate user (Eve) capable of eavesdropping on Alice-Bob's transmissions. The links between Alice and Bob (legitimate link) and between Alice and Eve (wiretap link) will be referred to as \emph{information links}, whereas the link between \ac{PB} and Alice will be denoted as the \emph{energy link}. For simplicity, yet without loss of generality, we consider that the time-sharing protocol proposed in \cite{Nasir2013} is used, where the energy and information phases have a duration of $T/2$ seconds, with $T$ being the time slot duration.

In Fig. \ref{PB_system}, $P_0$ is the available power at Alice's transmitter, where the function $g(\cdot)$ indicates the EH behavior, which in general will not be linear. Similarly, $P_I$ is the available power at the input of the EH (Alice), which can be expressed as 
\begin{equation}
	P_I = P_{T} L_P {d_{P}}^{-\alpha} |h_P|^2,
\end{equation}
where $P_T$ is the transmitted power by the \ac{PB}, $L_P$ incorporates the gains of the transmit and receive antennas and frequency-dependent propagation losses, $d_P$ is the distance between the \ac{PB} and Alice and $\alpha$ is the path loss exponent. 

The channels $h_P$, $h_B$ and $h_E$ are in charge of incorporating the random effects of multipath fading over the transmitted signal. Without loss of generality, we consider normalized fading channel coefficients so that $\mathbb{E}{\{|h_P|^2\}}=\mathbb{E}{\{|h_B|^2\}}=\mathbb{E}{\{|h_E|^2\}}=1$, where $\mathbb{E}\{\cdot\}$ is the expectation operator. We also assume that channels are quasi-static fading channels, (i.e. constant during the transmission of given codeword, yet independent between codewords unless otherwise stated), and the \ac{AWGN} also affects the communication through information links.

Now, depending on the scenario under consideration, channel correlation can be incorporated in different ways. In this paper, we will consider two situations: \emph{(i)} the energy and \emph{legitimate} links are correlated; and \emph{(ii)} the energy and \emph{wiretap} links are correlated. The former situation corresponds to the case on which the \ac{PB} also plays the role of Bob, and Alice uses the energy obtained from the \ac{PB} to report back some information to him. The latter scenario represents the case of an untrusted \ac{PB}, so Alice wants that the information transmitted to an external legitimate user Bob remains confidential to the \ac{PB}, which now acts as Eve.

The instantaneous \ac{SNR} at both receiver ends will be given by: 
%	\gamma_B= &\frac g {P_{T} L_B(d_P d_B)^{{{-\alpha}}}}{N_0} \left | {h_P} \right |^2 \left | {h_B} \right |^2\\ \notag
%	=&\overline{\gamma}_B^{I} \mathbb{E}\{\left | {h_P} \right |^2 \left | {h_B} \right |^2\} \left | {h_P} \right |^2 \left | {h_B} \right |^2,
\begin{align}
	\label{eq:SNR1}
	\gamma_B= &\, \frac {P_0 L_B {d_{B}}^{{-\alpha}} }{N_0} \left | {h_B} \right |^2 \\
	\gamma_E =& \frac {P_0 L_E {d_{E}}^{{-\alpha}} }{N_0}\left | {h_E} \right |^2%\notag\\ 
%	= & \,\overline{\gamma}_B^{I} \mathbb{E}\{\left | {h_P} \right |^2 \left | {h_B} \right |^2\} \left | {h_P} \right |^2 \left | {h_B} \right |^2,
\end{align}
%\mathbb{E}\{\left | {h_P} \right |^2 \left | {h_B} \right |^2\}
where %$\overline{\gamma}_B^{I} = \frac {P_{PB} \eta L_B(d_P d_B)^{{{-\alpha}}}}{N_0}$
%$\overline{\gamma}_B^{I} = {P_0 L_B {d_{B}}^{{-\alpha}}}/{N_0}$ and 
$N_0$ is the noise power, $P_0 = g (P_{T} L_P {d_{P}}^{{-\alpha}}\left | {h_P} \right |^2)$, $d_B$ and $d_E$ are the distances between Alice-Bob and Alice-Eve, respectively, and $\{L_B,L_E\}$ are defined in a similar way as $L_P$. 

The theoretical formulations in this work will be valid for an arbitrary choice of fading distributions. However, because the links in \ac{WPC} are of inherent \ac{LOS} nature, we will specifically consider that the different fading channel coefficients are modeled by the Rician distribution, which is characterized by the parameter $K$ defined as the ratio between the \ac{LOS} and non-\ac{LOS} powers. Now, denoting as $h_I$ the information link (either legitimate or wiretap) that is correlated with the energy link, we have that the normalized fading channels are expressed as
\begin{equation}
\label{eq4}
	h_P=\sqrt{\frac{K}{K+1}} + \sqrt{\frac{1}{K+1}} \, z_{1},
\end{equation}
\begin{equation}
\label{eq5}
	h_I=\sqrt{\frac{K}{K+1}} + \sqrt{\frac{1}{K+1}} \left(\rho \, z_{1} + \sqrt{1-\rho^2} \, z_{2}\right), 
\end{equation}
with $\rho=\frac{{\rm{cov}}(h_P,h_I)}{\sqrt{{\rm{var}}(h_P){\rm{var}}(h_I)}}$ being the correlation coefficient, and $z_1$ and $z_2$ being two ancillary complex normal RVs. Note that for $K=0$, the scenario reduces to the case of Rayleigh fading.

\section{Effect of Correlation in System's Capacity}
\label{CorrelationWPC}
\subsection{Preliminary definitions}
In order to understand the effect of correlation in the investigated set-up, we first will consider an eavesdropper-free scenario. In this situation, the receive SNR will be given by \eqref{eq:SNR1}. For simplicity, and because of the inherent complexity associated to considering correlation, we will assume that the non-linear \ac{EH} operates in the linear region so that the saturation effect associated to its operation can be neglected. Thus, our results can be seen as an upper bound on the achievable performance. %In our simulations, we observed that incorporating the \ac{EH} non-linearity causes an additional degradation on the receive SNR which is virtually agnostic to link correlation. Hence, both effects (correlation and non-linearity) can be studied separately.

Under these premises, we have that $P_O\approx\eta P_I$, with $\eta$ being the \ac{EH} efficiency. So, we can express ${\gamma}_B$ as
\begin{align}
\label{eq6}
	\gamma_B= &\, \underbrace{\frac {P_{T} L_P {d_{P}}^{{-\alpha}}\eta L_B {d_{B}}^{{-\alpha}} }{N_0}}_{\overline\gamma_B^I=P_T/N_E} \underbrace{\left | {h_P} \right |^2}_{x}\underbrace{\left | {h_B} \right |^2}_{y}.
\end{align}

For convenience of discussion, we define the parameter $\overline\gamma_B^I$ as the ratio between the \ac{PB} transmit power and a constant term $N_E$, which can be regarded as the system's noise referred to the \ac{PB} output. We note that $\overline\gamma_B^I$ reduces to the average SNR at the receiver side only in the absence of correlation. In such case, we have $\mathbb{E}\{ x y \}=\mathbb{E}\{ x\}\mathbb{E}\{y\}=1$ because of the definition of normalized channel gains. In the general case of correlation between the energy and information links, we have that $\mathbb{E}\{ x y\}>1$. 

\subsection{Performance analysis}
In order to analyze the effect of correlation, we use two classical performance metrics for benchmarking: average capacity and outage probability. The average capacity per bandwidth unit is defined as 
\begin{equation}
\label{eq7}
\overline{C}\triangleq\mathbb{E}\left\{\log_2(1+\gamma_B)\right\}=\int_0^{\infty}\log_2(1+\gamma)f_{\gamma_B}(\gamma)d\gamma,
\end{equation}
whereas the \ac{OP} is defined as the probability that the instantaneous SNR (or equivalently, the instantaneous capacity) falls below a predefined threshold, i.e.
\begin{align}
{\rm{OP}}=\Pr\left\{\log_2(1+\gamma_B)<R_{\rm th}\right\}=\Pr\left\{\gamma_B<\gamma^{\rm{th}}\right\},
\end{align}
with $\gamma^{\rm{th}}=2^{R_{\rm th}}-1$. 

We will now describe how these performance metrics can be derived from the joint distribution of $x=\left | {h_P} \right |^2$ and $y=\left | {h_B} \right |^2$. Even though the correct derivation only requires the use of standard techniques of transformation of random variables (e.g. see \cite[ch. 6]{Papoulis2002}), the complicated form of the bivariate distributions of $x$ and $y$ for most fading distributions makes it hard to evaluate the performance in a simple form. Starting from the bivariate PDF of $x$ and $y$, we can obtain the PDF of the RV $u=x\cdot y$ as \cite[eq. 6.74]{Papoulis2002}
\begin{equation}
f_{u}(u)=\int_0^{\infty}\frac{1}{t}f_{x,y}\left(t,\frac{u}{t}\right)dt,
\end{equation}
and then we can obtain 
\begin{equation}
\label{eqCint}
\overline{C}=\int_0^{\infty}\log_2(1+\overline\gamma_{B}^{I}\cdot u)f_{u}(u)du
\end{equation} and 
\begin{equation}
\label{eqOPint}
{\rm{OP}}=F_{u}\left(\gamma^{\rm{th}}/\overline\gamma_{B}^{I}\right)=\int_0^{\gamma^{\rm{th}}/\gamma_{B}^{I}}f_u(u)du,
\end{equation}
where $F_u(u)$ is the CDF of $u$. For the general case of Rice fading the bivariate distribution has a rather intricate form, which is here reproduced for the readers' convenience \cite{Simon2007}:
\begin{align}
f_{x,y}(x,y)&=\frac{(K+1)^2}{1-\rho^2} e^{-\frac{K+1}{1-\rho^2}(x+y)}e^{-\frac{2K}{1+\rho}} \times\nonumber\\ &\sum_{k=0}^{\infty}\varepsilon_k I_{k}\left(\alpha\sqrt{xy}\right)I_{k}\left(\beta\sqrt{x}\right)I_{k}\left(\beta\sqrt{y}\right),
\end{align}
with $\alpha=\frac{2\rho(K+1)}{1-\rho^2}$, $\beta=\frac{2}{1+\rho}\sqrt{K(K+1)}$, $I_{k}(\cdot)$ the modified Bessel function of the first kind and order $k$, and $\varepsilon_k$ being the Neumann constant $\varepsilon_0=1$ and $\varepsilon_{k>0}=2$. Unfortunately, the derivation of tractable performance metrics for the correlated Rician case is not possible, and double integral expressions are required for the evaluation of the capacity and outage probability. However, they can be easily computed with standard numerical integration routines (e.g. using \texttt{integral} in \MATLAB).

From the definitions in \eqref{eq4}, \eqref{eq5} and \eqref{eq6}, and after some manipulations we can see that
\begin{equation}
\label{eqmoments}
\mathbb{E}\{\gamma_B\}=\overline\gamma_B=\overline\gamma_B^I\underbrace{\frac{K^2+2K(1+\rho)+1+\rho^2}{(K+1)^2}}_{\Delta_{\rm PO}}.
\end{equation}
which for Rayleigh fading reduces to $\mathbb{E}\{\gamma_B\}=\overline\gamma_B^I(1+\rho^2)$. A first remark is in order at this point: we can see that for a fixed transmit power $P_T$, the average SNR at the receiver end is increased due to correlation, i.e., $\overline\gamma_B\geq\overline\gamma_B^I$, or in other words $\Delta_{\rm PO}\geq 1$. While this may seem beneficial from a system design perspective, it also increases the variance of the equivalent composite channel, which implies a larger fading severity.

We can use the previous results to obtain an approximate expression for the average capacity in the high \ac{SNR} regime, in the form
\begin{equation}
\label{eqC}
\overline{C}\approx \log_2(\overline\gamma_B)-t.
\end{equation}
The term $t$ can be regarded as a capacity loss with respect to the \ac{AWGN} case (i.e., the absence of fading), and can be computed from the expression of the moments of $\gamma_{B}=\overline\gamma^I_B\cdot u$ following the procedure described in \cite{Yilmaz2012}. In our case, obtaining a closed-form expression for the moments can be complicated due to the intractable form of the \ac{PDF} of $\gamma_B$. However, in the case of independent links, the moments of $\gamma_B$ can be computed from the product of the moments of independent Rician \acp{RV}. Thus, the capacity loss in the Rician product channel is twice the capacity loss of the single Rician link (e.g., given in \cite{Rao2015}), i.e.
\begin{equation}
\label{trho0}
t_{(\rho=0)}=2\left(\log_2\left(\frac{K+1}{K}\right)-\log_2(e){\rm E}_1(K)\right),
\end{equation}
where ${\rm E}_1(\cdot)$ is the exponential integral function. Now, the scale factor $\Delta_{\rm PO}$ is translated into an additional capacity penalty for a fixed $\overline{\gamma}_B$, as in \cite{Matez2020}. Thus, we can express from \eqref{eqmoments}:
\begin{align}
t=&t_{(\rho=0)}+t(\rho)=t_{(\rho=0)}+\log_2(\Delta_{PO})\nonumber\\
=&t_{(\rho=0)}+\log_2\left(\frac{K^2+2K(1+\rho)+1+\rho^2}{(K+1)^2}\right)
\label{eqt}
\end{align}
We can see that the capacity loss $t$ grows as $K$ is reduced or as $\rho$ is increased, so that system performance is expected to be degraded with correlation and the absence of \ac{LOS}. We note that \eqref{eqt} reduces to the results recently given in \cite{Matez2020} for the Rayleigh case, since $\lim_{K\rightarrow0} \left\{\log\left(\frac{K+1}{K}\right)-{\rm E}_1(K) \right\}= \gamma_e$, where $\gamma_e$ is the Euler-Mascheroni constant and $\log$ is the natural logarithm.

Due to the unwieldy nature of the distribution of the product of correlated Rician RVs, we will now pay attention to the simplified case of Rayleigh fading. In order to better understand the effect of correlation on fading severity, we can also resort to the \ac{AoF} parameter, which for the case of a correlated Rayleigh product channel can be computed as
\begin{equation}
{\rm{AoF}}=\frac{\mathbb{E}[\gamma_B^2]}{\overline{\gamma}_B^2}-1=\frac{4(1+p(4+p))}{(1+p)^2}-1,
\end{equation}
% a=4*(1+p*(4+p))./(1+p).^2-1;
where $p=\rho^2$ for the sake of shorthand notation, and we used the definition of the \ac{PDF} given in \cite[eq. 6.55]{Simon2007} as:
\begin{equation}
\label{eqPDF}
f_{\gamma_B}(\gamma)=\tfrac{2}{\overline\gamma_B}\tfrac{1+p}{1-p}I_0\left(\tfrac{2}{1-p}\sqrt{\tfrac{p\gamma(1+p)}{\overline\gamma_B}}\right)K_0\left(\tfrac{2}{1-p}\sqrt{\tfrac{\gamma(1+p)}{\overline\gamma_B}}\right),
\end{equation}
where $K_\nu(\cdot)$ is $\nu$-th order modified Bessel functions of the second kind, and $\mathbb{E}\{\overline\gamma_B\}=\overline\gamma_B$ denotes the average \ac{SNR} at the receiver. As predicted by the asymptotic capacity results, we see that as $p$ grows the \ac{AoF} increases, so that fading severity increases with correlation and ultimately degrades system performance.

The \ac{OP} can be directly computed by evaluating the \ac{CDF} of $\gamma_B$, i.e., ${\rm OP}=F_{\gamma_B}\left(\gamma^{\rm th}\right)$, with
\begin{align}
\label{eqCDF1}
F_{\gamma_B}(\gamma)=1-a\sqrt{\gamma}&\left[I_0(a\sqrt{p\gamma})K_1(a\sqrt{\gamma})+\right.\notag\\&\left.\sqrt{p}I_1(a\sqrt{p\gamma})K_0(a\sqrt{\gamma})\right],
\end{align}
which is obtained by manipulating the expression in \cite[eq. 6]{Bithas2007}, with $a=\frac{2}{1-p}\sqrt{\frac{1+p}{\overline\gamma_B}}$.

\subsection{Numerical results}
\begin{figure}[t]
	\centering
		\includegraphics{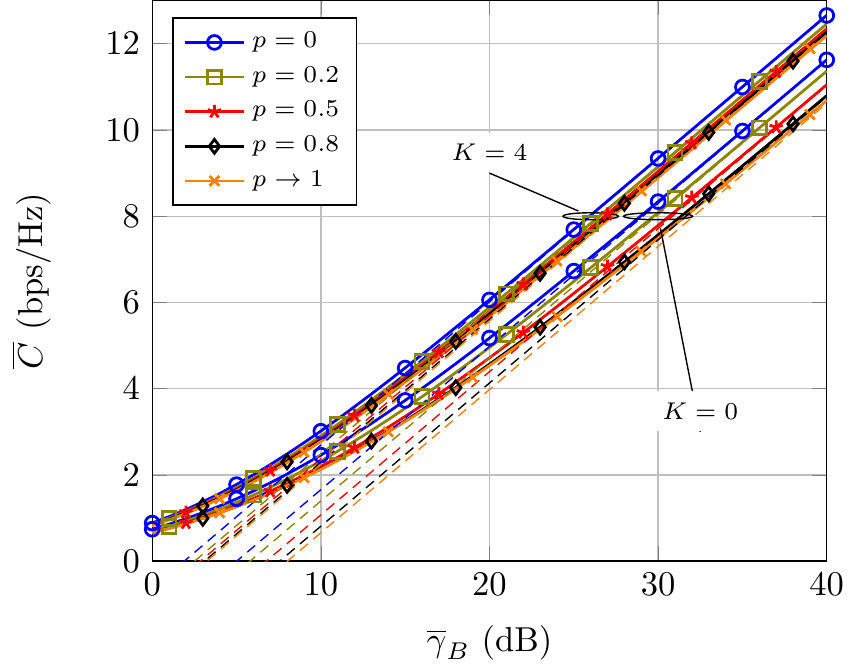}
	\caption{Average capacity $\overline{C}$ vs. $\overline\gamma_B$, for different values of the correlation coefficient through $p=\rho^2$ and \ac{LOS} conditions through $K$. Solid lines correspond to theoretical expressions using \eqref{eqCint} and \eqref{eq7}. Dashed lines correspond to the asymptotic results using \eqref{eqC}. Markers correspond to MC simulations.} 
	\label{fig:2}
\end{figure}

\begin{figure}[t]
	\centering
		\includegraphics{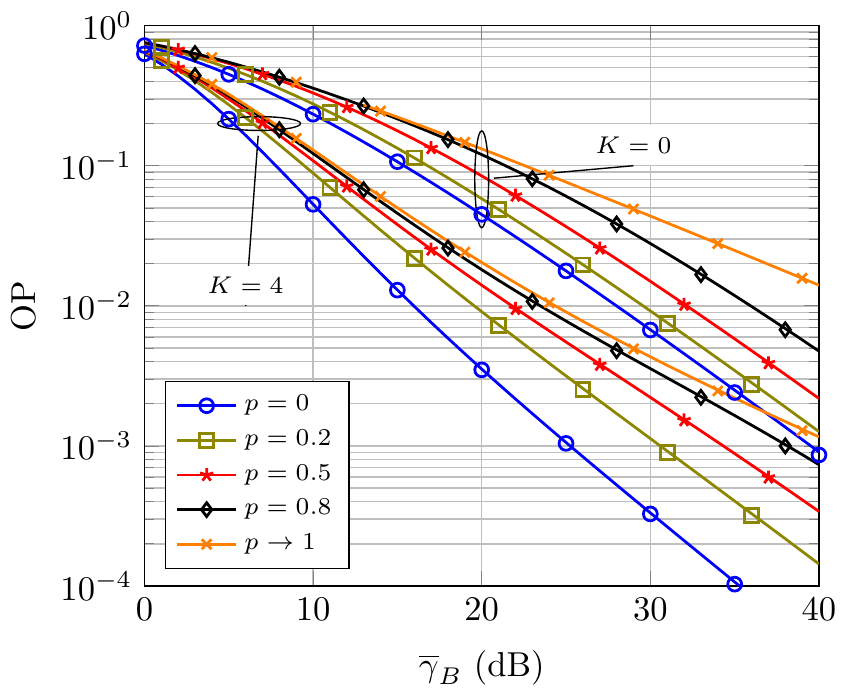}
		\caption{Outage probability vs. $\overline\gamma_B$, for different values of the correlation coefficient through $p=\rho^2$ and \ac{LOS} conditions through $K$. Solid lines correspond to theoretical expressions using \eqref{eqOPint} and \eqref{eqCDF1}. Markers correspond to MC simulations.} 
	\label{fig:3}
\end{figure}

In Figs. \ref{fig:2} and \ref{fig:3}, we evaluate the average capacity and the \ac{OP} as a function of the average received \ac{SNR}, for different values of the correlation coefficient through $p=\rho^2$. For the \ac{OP} evaluation, we set $R_{\rm th}=1$ bps/Hz. The infinite series expression required for the evaluation of the performance metrics in the Rician case has been truncated to 10 terms, which is sufficient for an excellent match with \ac{MC} simulations. We see that in all instances, correlation degrades system performance when compared to the case of independent energy and information links. However, the effect of link correlation is reduced as the \ac{LOS} condition is increased. i.e. as $K$ grows. As discussed in \cite{Matez2020}, the average capacity loss due to correlation implies that a power offset is required in the presence of correlation in order to obtain the same capacity. Interestingly, such power offset $\Delta_{\rm PO}^{\rm dB}=10\log_{10}(\Delta_{\rm PO})$ dB is compensated by the fact that the required transmit power $P_T$ is reduced in the event of correlation by a factor of $\mathbb{E}\{x y\}=\Delta_{\rm PO}$. Hence, the performance loss due to correlation is compensated by the increase in average SNR for a fixed power budget. However, this does not seem the case for the \ac{OP} (at least for operational ranges of low \ac{OP}s), for which the power offset required to achieve a target \ac{OP} is much larger, as observed in Fig. \ref{fig:3}.

\section{Effect of Correlation on Secrecy Capacity}
\subsection{Preliminary definitions}
We will now analyze the effect of channel correlation on the physical layer security performance of the system model under consideration. Such secure performance is quantified through the secrecy capacity $C_{\rm S}$, defined as \cite{Bloch2008}:
\begin{equation}
\label{eqISC}
C_{\rm S}\triangleq \max\{\log_2(1+\gamma_B)-\log_2(1+\gamma_E),0\}.
\end{equation}

Depending on the assumptions taken regarding availability of \ac{CSI} at the different agents, the \ac{OP} of secrecy capacity or the average secrecy capacity are the usual metrics to characterize the performance in this context, namely
\begin{equation}
{\rm OP_{SC}}\triangleq \Pr\{\log_2(1+\gamma_B)-\log_2(1+\gamma_E)<R_{\rm th}\}.
\end{equation}
\begin{equation}
\overline C_{\rm S}\triangleq \underset{\gamma_B>\gamma_E}{\mathbb{E}}\{\log_2(1+\gamma_B)-\log_2(1+\gamma_E)\},
\end{equation}

Note that in order to analytically calculate these performance metrics, the joint distribution of $\gamma_B$ and $\gamma_E$ is required. Depending on the way that channel correlation is incorporated between the energy and information links, such distribution will take different forms. The two scenarios described in Section II, namely (\emph{i}) correlation between the energy and legitimate information link and (\emph{ii}) correlation between the energy and wiretap link, are tackled in the next subsections. %Due to the rather intricate nature of the resulting distributions, we resort to simulations in this section of the paper.

\label{CorrelationCS}
\subsection{Correlation between WPT and Bob's links}
\label{S1}

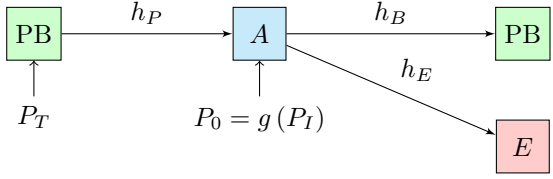
\begin{figure}[t]
\centering
\begin{tikzpicture}[node distance=3.5cm,auto,>=latex']
    \node [int, pin={[init]below:$P_0=g\left(P_{I}\right)$}] (a) {$A$};
    \node [int_green, pin={[init]below:$P_{T}$}, left of=a,node distance=3cm] (b) {$\rm{PB}$};
    \node [int_green] (c) [right of=a] {$\rm{PB}$};
    \node [int_red] (e) [below of=c, node distance=1.5cm] {$E$};
    \node [coordinate] (end) [right of=c, node distance=2cm]{};
    %\path[->] (b) edge node {$a$} (a);
    \path[->] (a) edge node {$h_B$} (c);
    %\draw[->] (c) edge node {$p$} (end) ;
    \path[->] (a) edge node {$h_E$} (e);
    \path[->] (b) edge node {$h_{P}$} (a);
\end{tikzpicture}
\caption{Scenario 1: Correlation between the energy and legitimate information links.}
\label{PB_system2}
\end{figure}

We first consider the case on which Bob plays the role of \ac{PB}, whereas Alice is the \ac{EH} device. Both agents wish to communicate in the presence of an external eavesdropper Eve, for which its wireless channel is statistically independent to the ones between Alice and Bob. Hence, the correlation parameter $\rho$ will capture the correlation between the energy and legitimate information links. This scenario, depicted in Fig. \ref{PB_system2}, can be regarded as an extension of that considered in the previous section, only that now adding Eve to the system under analysis.

We must note that while the average SNR at Bob may be increased due to correlation for a fixed $P_T$, this effect does not take place on the average SNR experienced by Eve. Similarly, if we set a fixed average SNR at Bob, the required transmit power by the \ac{PB} is reduced because of correlation by a factor $\Delta_{\rm PO}=\mathbb{E}\{\left | {h_P} \right |^2\left | {h_B} \right |^2\}$. Hence, \emph{this causes a reduction on the average SNR at Eve by the same factor}, {which seems beneficial from a physical layer security perspective}. In any case, it is clear that despite correlation may not be present between the energy and eavesdropper's links, it clearly influences the relation between the average SNRs at the legitimate and eavesdropper's ends. This effect has been overlooked in the literature, and not properly introduced to the best of our knowledge, e.g., see \cite{Zhang19b} for an $\rm OP_{SC}$ analysis in the presence of correlation in the context of backscatter communications. 

%\subsubsection{Performance analysis}
The derivation of analytical expressions for the secrecy performance metrics is even more complicated than in the scenario tackled in Section III. For this reason, we aim to describe how the evaluation of these performance metrics can be carried out in a general form. Let us express the instantaneous secrecy capacity in \eqref{eqCint} in a more convenient form:
\begin{align}
\label{eqISC2}
C_{\rm S}&= \max\left\{\log_2\left(\frac{1+\overline\gamma_B^I x\cdot y}{1+\overline\gamma_E x\cdot z}\right),0\right\}.
\end{align}
 where $x=\left | {h_P} \right |^2$, $y=\left|{h_B} \right |^2$ and $z=\left | {h_E} \right |^2$. Now, defining $u=x\cdot y$ and $v=x\cdot z$, we can obtain the joint distribution of $u$ and $v$ from the joint distribution of $x$, $y$, and $z$. In the scenario under consideration, we have that
 \begin{align}
 f_{x,y,z}(x,y,z)= f_{x,y}(x,y)\cdot f_{z}(z),
 \end{align}
 because of the independence of the eavesdropper's link. Defining the ancillary \ac{RV} $w=z$, we can compute the joint distribution of $u$ and $v$ using standard techniques of transformation of \acp{RV} as
 \begin{align}
 f_{u,v}(u,v)=&\int_{0}^{\infty} f_{u,v,w}(u,v,w)dw,\nonumber\\
 =&\int_{0}^{\infty} \frac{f_{x,y}(v/w,uw/v)\cdot f_{z}(w)}{\left|\frac{\partial (u,v,w)}{\partial(x,y,z)}\right|}dw,\nonumber\\
 =&\int_{0}^{\infty} \frac{1}{v}f_{x,y}(v/w,uw/v)\cdot f_{z}(w)dw,\label{eqjoint}
 \end{align}
where ${\left|\frac{\partial (u,v,w)}{\partial(x,y,z)}\right|}=v$ is the Jacobian of the \ac{RV} transformation. Now, \eqref{eqjoint} can be used to derive the secrecy performance metrics as
 \begin{equation}
 \label{eqASC}
\overline C_{\rm S}=\int_{0}^{\infty}\int_{0}^{u\overline\gamma^I_B/\overline\gamma_E}\log_2\left(\frac{1+\overline\gamma_B^I u}{1+\overline\gamma_E v}\right)f_{u,v}(u,v)dvdu
\end{equation}
\begin{equation}
\label{eqopsc}
{\rm OP_{SC}}=\int_{0}^{\infty}\int_{\max\{0,bu-c\}}^{\infty}f_{u,v}(u,v)dvdu,
\end{equation}
with $b=\overline\gamma_B^I/\left(2^{R_{\rm th}}\overline\gamma_E\right)$ and $c=\left(2^{R_{\rm th}}-1\right)/\left(2^{R_{\rm th}}\overline\gamma_E\right)$

It is evident that obtaining analytical expressions for these metrics in the general case of Rician fading seems unlikely. In the simplified case of Rayleigh fading, it is possible to obtain a closed form expression of $ f_{u,v}(u,v)$ by integrating the bivariate distribution of two correlated Rayleigh RVs \cite{Simon2007} as in \eqref{eqjoint}, yielding
\begin{align}
\label{eqfuv}
 f_{u,v}(u,v)=\tfrac{\alpha}{\sqrt{u+2v/\alpha}}I_0\left(\alpha\sqrt{u p}\right)K_1\left(\alpha\sqrt{u+2v/\alpha}\right)
 \end{align}
where again $p=\rho^2$ and $\alpha=2/(1-p)$. Plugging \eqref{eqfuv} in \eqref{eqopsc}, we obtain a single integral expression for the \ac{OP} of secrecy capacity by direct integration as
\begin{equation}
\label{eqopsc2}
{\rm OP_{SC}}=\int_{0}^{\infty}g(u) du,
\end{equation}
with
\begin{equation}
\label{eqopsc3}
g(u)=\alpha I_0\left(\alpha\sqrt{u p}\right) K_0\left(\alpha\sqrt{u+2\max\{0,bu-c\}/\alpha}\right).
\end{equation}

The average secrecy capacity is far more intricate to be computed analytically, although once again the powerful numerical integration routines included in \MATLAB\, or \MATHE\, can be used to evaluate this metric. However, we can resort to an asymptotic approximation in the high \ac{SNR} regime inspired by the analysis in \cite{Moualeu2019} as:
\begin{align}
\label{asc1}
\overline{C}_S\approx& \overline{C}_B-\overline{C}_E,\\
\approx& \log_2(\overline\gamma_B)-t-\overline{C}_E.
\label{asc2}
\end{align}
In this case, $\overline{C}_B$ (or equivalently, $t$) can be computed following the procedure detailed in Section III, while $\overline{C}_E$ is the capacity of an uncorrelated product channel.

\subsection{Correlation between WPT and Eve's links}
\label{S2}
We now analyze the case on which the \ac{EH} device, Alice, wishes to communicate to a legitimate device Bob while keeping the information confidential from the \ac{PB} that conveys energy to her. We refer to this scenario as \emph{secure communication in the presence of an untrusted \ac{PB}}. We note that this scenario on which the \ac{PB} plays the role of Eve has not been addressed in the literature to the best of our knowledge.

\begin{figure}[t]
\centering
\begin{tikzpicture}[node distance=3.5cm,auto,>=latex']
    \node [int, pin={[init]below:$P_0=g\left(P_{I}\right)$}] (a) {$A$};
    \node [int_red, pin={[init]below:$P_{T}$}, left of=a,node distance=3cm] (b) {$\rm{PB}$};
    \node [int_green] (c) [right of=a] {$B$};
    \node [int_red] (e) [below of=c, node distance=1.5cm] {$\rm{PB}$};
    \node [coordinate] (end) [right of=c, node distance=2cm]{};
    %\path[->] (b) edge node {$a$} (a);
    \path[->] (a) edge node {$h_B$} (c);
    %\draw[->] (c) edge node {$p$} (end) ;
    \path[->] (a) edge node {$h_E$} (e);
    \path[->] (b) edge node {$h_{P}$} (a);
\end{tikzpicture}
\caption{Scenario 2: Correlation between the energy and wiretap links.}
\label{PB_system3}
\end{figure}
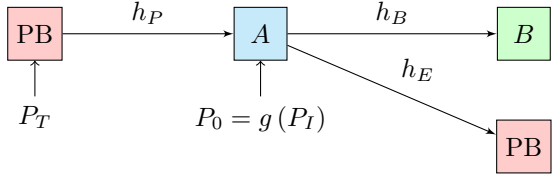

The differences between the scenario under consideration and that in Section \ref{S1} are given as follows: in this configuration, the energy and wiretap links are correlated and identically distributed, and in general will exhibit a \ac{LOS} condition. Now, the legitimate link is agnostic to correlation, which implies that the average SNR at Bob's receiver is not modified because of correlation for a given $P_T$. However, in this case correlation affects the composite energy-wiretap link in two ways: for a fixed $P_T$, the average SNR received at the \ac{PB} (i.e., Eve) is now increased by a factor $\Delta_{\rm PO}=\mathbb{E}\{\left | {h_P} \right |^2\left | {h_E} \right |^2\}$,  i.e. $\overline\gamma_E=\overline\gamma_E^I \Delta_{\rm PO}$, where $\overline\gamma_E^I $ is the average \ac{SNR} at Eve in the case of independent links. Besides, correlation increases fading severity for such composite link, which also impacts capacity of the wiretap link.

Let us follow a similar approach as in the previous subsection to describe the general way to computing the secrecy performance metrics in this scenario. Starting from the expression in \eqref{eqISC2}, and using the same definitions for $x$, $y$, and $z$, we now have that
 \begin{align}
 f_{x,y,z}(x,y,z)= f_{x,z}(x,z)\cdot f_{y}(y),
 \end{align}
 because of the independence of the legitimate link. We now define the ancillary \ac{RV} $w=y$, and then compute the joint distribution of $u$ and $v$ as
 \begin{align}
 f_{u,v}(u,v)=&\int_{0}^{\infty} f_{u,v,w}(u,v,w)dw,\nonumber\\
 =&\int_{0}^{\infty} \frac{f_{x,z}(u/w,vw/u)\cdot f_{y}(w)}{\left|\frac{\partial (u,v,w)}{\partial(x,y,z)}\right|}dw,\nonumber\\
 =&\int_{0}^{\infty} \frac{1}{u}f_{x,z}(u/w,vw/u)\cdot f_{y}(w)dw.\label{eqjoint2}
 \end{align}
In this case \eqref{eqjoint2} can be used to evaluate the secrecy performance metrics as
 \begin{equation}
 \label{eqASC2}
\overline C_{\rm S}=\int_{0}^{\infty}\int_{0}^{u\overline\gamma_B/\overline\gamma_E^I}\log_2\left(\frac{1+\overline\gamma_B u}{1+\overline\gamma_E^I v}\right)f_{u,v}(u,v)dvdu
\end{equation}
\begin{equation}
\label{eqopscB}
{\rm OP_{SC}}=\int_{0}^{\infty}\int_{0}^{\beta v+\psi}f_{u,v}(u,v)dudv,
\end{equation}
with $\beta=\left(2^{R_{\rm th}}\overline\gamma_E^I\right)/\overline\gamma_B$ and $\psi=\left(2^R_{\rm th}-1\right)/\overline\gamma_B$. 

The derivation of analytical expressions for these metrics seems challenging, if not impossible, for the general case of Rician fading. However, reasonably tractable expressions can be obtained in the Rayleigh case for the \ac{OP} of secrecy capacity. Following similar steps as those described in Section \ref{S1} we can obtain
\begin{equation}
\label{eqopscB2}
{\rm OP_{SC}}=1-\int_{0}^{\infty}g(v) dv,
\end{equation}
with
\begin{equation}
\label{eqopscB3}
g(v)=\alpha I_0\left(\alpha\sqrt{v p}\right) K_0\left(\alpha\sqrt{2\psi/\alpha+v(2\beta\alpha+1)}\right).
\end{equation}

The asymptotic secrecy capacity can also be approximated as
\begin{align}
\label{asc3}
\overline{C}_S\approx& \overline{C}_B-\overline{C}_E,\\
\approx& \log_2(\overline\gamma_B)-t_{(\rho=0)}-\overline{C}_E.
\label{asc4}
\end{align}
where now $\overline{C}_B$ is the capacity of an uncorrelated product channel, the parameter $t_{(\rho=0)}$ is given in \eqref{trho0} and $\overline{C}_E$ can be calculated be computed as in Section III.

%so we use the same value of the Rician $K$ parameter to indicate their \ac{LOS} condition. We also assume that there is no \ac{LOS} between Alice and Bob. Now, the legitimate link is agnostic to correlation, as the average SNR at Bob's receiver is not modified %\footnote{Indeed, Alice can design the amount of reflected power used for communication, although such possibility is not analyzed here for the sake of simplicity.} 
 %because of correlation for a given $P_T$. 

\subsection{Numerical evaluation}
\begin{figure}[t]
	\centering
		\includegraphics{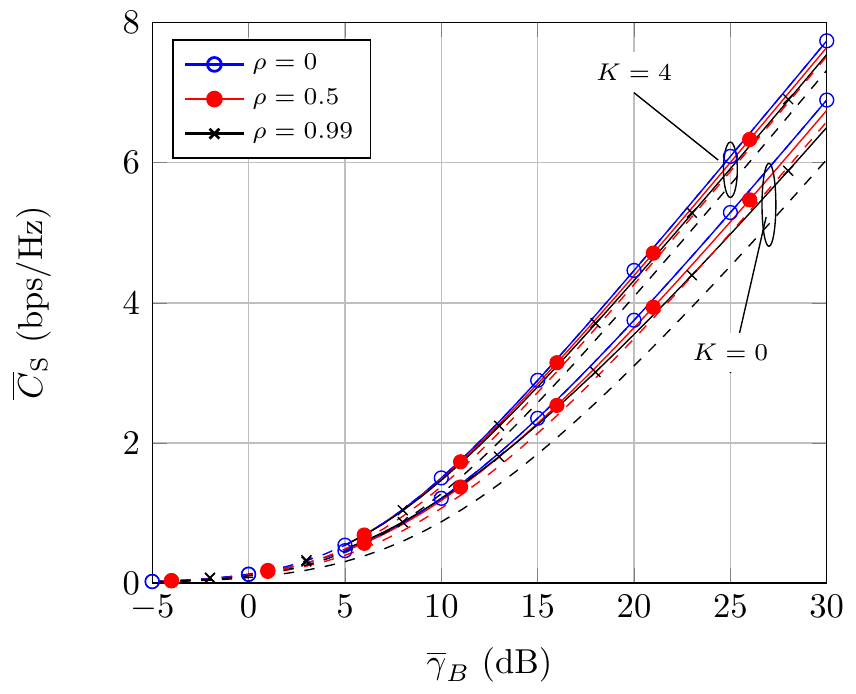}
	\caption{Average secrecy capacity $\overline{C}_{\rm S}$ vs. $\overline\gamma_B$, for different values of the correlation coefficient $\rho$ and \ac{LOS} conditions. Solid lines correspond to the case on which $\overline\gamma_E=5$dB for $\rho=0$, and then reduced by a factor $\Delta_{\rm PO}$. Dashed lines indicate the incorrect case where $\overline\gamma_E=5$ dB regardless of $\rho$. Markers correspond to \ac{MC} simulations, and the theoretical expressions in \eqref{eqASC} have been used to generate the dashed and solid line plots.} 
	\label{fig:4}
\end{figure}

%Hence, we will analyze its impact in the following set of figures.
%
We will now move to the numerical evaluation of the performance metrics in the aforementioned scenarios. Let us first begin with the case (\emph{i}) on which the energy and legitimate information links are correlated. In Fig. \ref{fig:4}, we represent the average secrecy capacity for a fixed average SNR at Bob, assuming different propagation conditions (\ac{LOS} and non-\ac{LOS}) and different values of $\rho$. %\footnote{For the sake of simplicity, we now use $\rho$ instead of the power correlation coefficient $p$ for comparison. Note that for a given $\rho$, the value of $p$ changes with $K$ for the case of Rician fading.}. 
 Because the energy and legitimate information links are identically distributed, we use the same value of the Rician $K$ parameter to quantify their \ac{LOS} condition. We consider that the eavesdropper's link is non-\ac{LOS} (i.e., Rayleigh), and set $\overline\gamma_E=5$dB for the case of independent energy and legitimate information links. Because we are comparing the average secrecy capacity for a target value of $\overline\gamma_B$, the transmit power $P_T$ is reduced due to correlation for a given set-up and the average SNR at the eavesdropper is reduced by a factor $\Delta_{\rm PO}$. This case, on which the \emph{true} average SNR at Eve is assumed, is represented by solid lines in the plot. For the sake of comparison, we also include with dashed lines the case on which $\overline\gamma_E$ is (incorrectly) kept invariant regardless of $\rho$ for a fixed $\overline\gamma_B$, i.e., the average SNR at the eavesdropper side is artificially larger than its true SNR for $\rho>0$. As in the previous section, the infinite series expression within the bivariate Rician distribution is truncated to 10 terms.

\begin{figure}[t]
	\centering
		\includegraphics{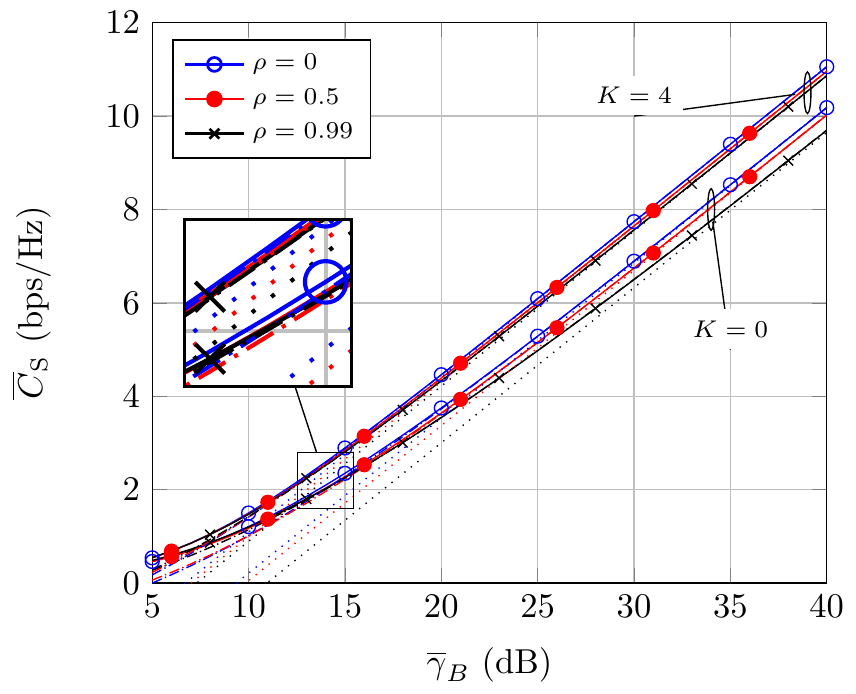}
	\caption{Exact and approximate average secrecy capacity $\overline{C}_{\rm S}$ vs. $\overline\gamma_B$, for different values of the correlation coefficient $\rho$ and \ac{LOS} conditions. For the eavesdropper's channel $\overline\gamma_E=5$dB for $\rho=0$, and then reduced by a factor $\Delta_{\rm PO}$. Solid lines indicate the exact theoretical expressions in \eqref{eqASC}, dash-dotted lines indicate the approximate expressions using \eqref{asc1} and dotted lines indicate the approximate expressions using \eqref{asc2}. Markers correspond to \ac{MC} simulations. } 
	\label{fig:4b}
\end{figure}

\begin{figure}[t]
	\centering
		\includegraphics{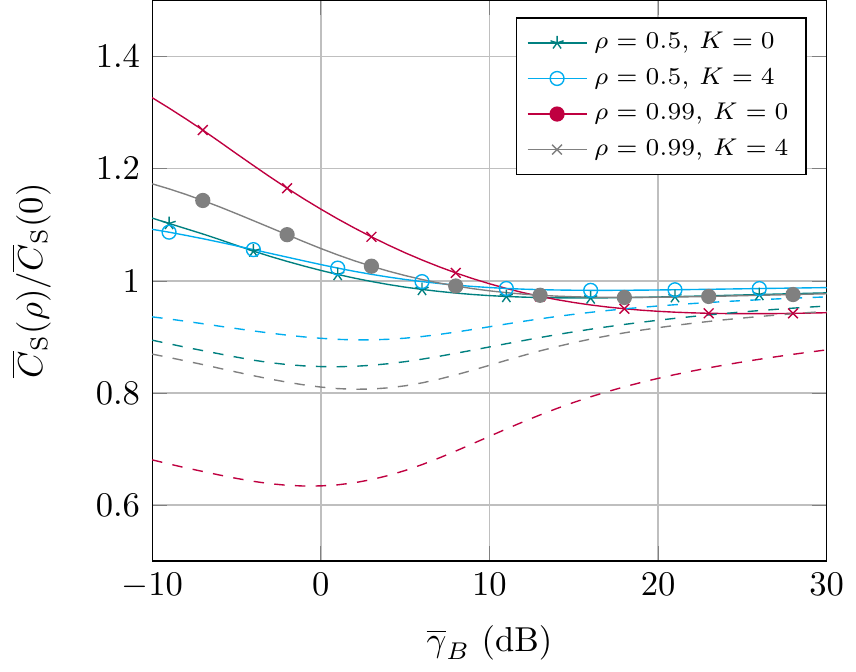}
	\caption{Average secrecy capacity $\overline{C}_{\rm S}$ normalized to that when $\rho=0$ vs. $\overline\gamma_B$, for different values of the correlation coefficient $\rho$ and \ac{LOS} conditions. Solid lines correspond to the case on which $\overline\gamma_E=5$dB for $\rho=0$, and then reduced by a factor $\Delta_{\rm PO}$. Dashed lines indicate the incorrect case where $\overline\gamma_E=5$ dB regardless of $\rho$. Markers correspond to \ac{MC} simulations, and the theoretical expressions in \eqref{eqASC} have been used to generate the dashed and solid line plots.} 
	\label{fig:5}
\end{figure}

We extract several important insights from the observation of Figs. \ref{fig:4} and \ref{fig:4}: \emph{(i)} for the investigated configuration, correlation is detrimental for secrecy capacity in the high SNR regime; \emph{(ii)} the asymptotic approximation given by \eqref{asc1} is very accurate in the medium and high \ac{SNR} regimes, whereas that obtained with \eqref{asc2} becomes tight only in the high \ac{SNR} regime; \emph{(iii)} the impact of correlation is lowered as the \ac{LOS} condition is increased; \emph{(iv)} neglecting the impact of correlation on the average SNR at Eve causes that the secrecy rate is underestimated compared to the \emph{true} SNR case. Importantly, because correlation increases fading severity of the legitimate link but also indirectly decreases the average SNR at Eve, it could be the case that the latter effect dominates over the former and correlation may be beneficial for capacity. In order to better visualize this, we represent in Fig. \ref{fig:5} the average secrecy capacity normalized to that obtained with $\rho=0$.

We see that as the average SNR at Bob is reduced, correlation turns out being beneficial for the average secrecy capacity, as we can obtain a larger secrecy rate than in the case of independent links. Such effect is intensified in the case of not having \ac{LOS} between the legitimate peers, although note that this corresponds to a \emph{relative} increase in capacity. Interestingly, when the \emph{true} average SNR at Eve is not considered, the secrecy rates seem always lower than those obtained with $\rho=0$.

\begin{figure}[t]
	\centering
		\includegraphics{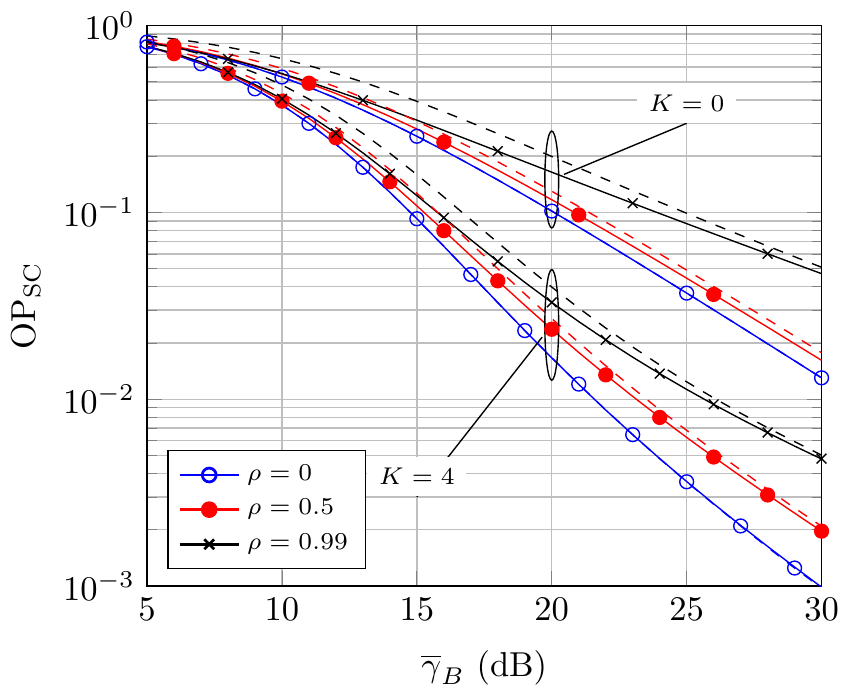}
	\caption{\ac{OP} of secrecy capacity $\overline{C}_{\rm S}$ vs. $\overline\gamma_B$, for different values of the correlation coefficient $\rho$ and \ac{LOS} conditions, with $R_{\rm th}=1$ bps/Hz.. Solid lines correspond to the case on which $\overline\gamma_E=5$dB for $\rho=0$, and then reduced by a factor $\Delta_{\rm PO}$. Dashed lines indicate the incorrect case where $\overline\gamma_E=5$ dB regardless of $\rho$. Markers correspond to \ac{MC} simulations, and the theoretical expressions in \eqref{eqopsc} and \eqref{eqopsc2} have been used to generate the dashed and solid line plots.} 
	\label{fig:6}
\end{figure}

In Fig. \ref{fig:6}, we now pay attention to the \ac{OP} of secrecy capacity, for the same set of parameters previously considered in Figs \ref{fig:4} and \ref{fig:5} and $R_{\rm th}=1$ bps/Hz. Similar observations can be made as those in the previous figures: correlation is detrimental for physical layer security in the operational range of OP values, although performance improves with $\ac{LOS}$ condition for the legitimate link. We also confirm that the impact of correlation in the \emph{true} average SNR at the eavesdropper's end is beneficial from a physical layer security perspective, compared to the erroneous situation on which correlation is neglected on $\overline\gamma_E$.

\begin{figure}[t]
	\centering
		\includegraphics{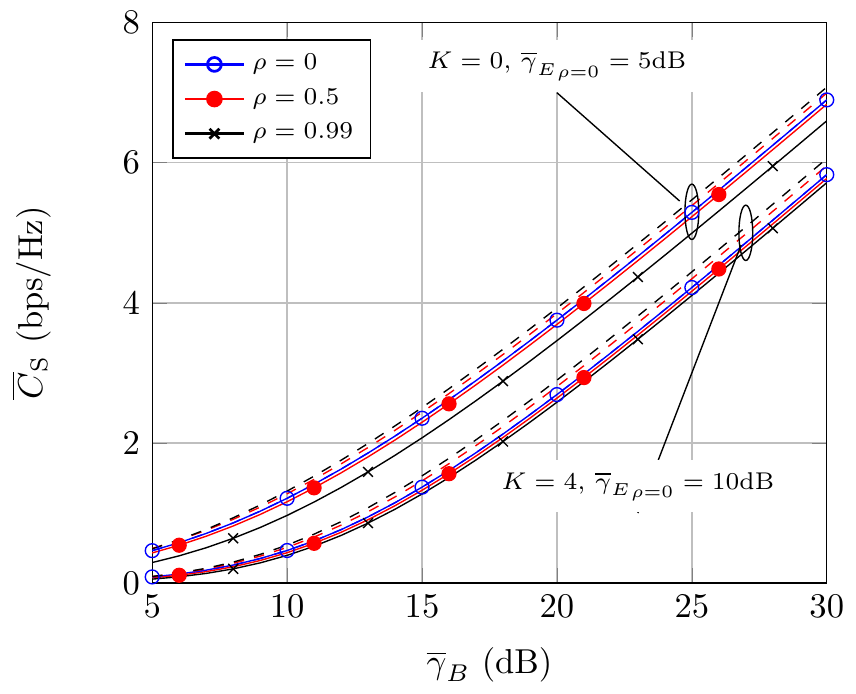}
	\caption{Average secrecy capacity $\overline{C}_{\rm S}$ vs. $\overline\gamma_B$, for different values of the correlation coefficient $\rho$ and \ac{LOS} conditions. Solid lines correspond to the case on which $\overline\gamma_E=\{5,10\}$dB for $\rho=0$, and then enlarged by a factor $\Delta_{\rm PO}$. Dashed lines indicate the incorrect case where $\overline\gamma_E=\{5,10\}$ dB regardless of $\rho$. Markers correspond to \ac{MC} simulations, and the theoretical expressions in \eqref{eqASC2} have been used to generate the dashed and solid line plots.} 
	\label{fig:7}
\end{figure}

Now, we will evaluate the secrecy performance of the scenario (\emph{ii}) that considers correlation between the energy and wiretap links. As in the previous configuration, we quantify the impact of correlation on physical layer security through two performance metrics: average secrecy capacity and \ac{OP} of secrecy capacity. In Fig. \ref{fig:7}, we evaluate the average secrecy capacity in \ac{LOS} and non-\ac{LOS} conditions, for different values of the correlation coefficient $\rho$. We assume that the link between Alice and Bob is of non-\ac{LOS} nature (i.e., subject to Rayleigh fading). Because the effect of \ac{LOS} in the energy and eavesdropper's links is not as important as in the scenario in subsection \ref{S1}, we use different values for the average SNR at the eavesdropper in order to improve the readability of the figure. We see that the average secrecy capacity is decreased as correlation grows, being this effect specially noteworthy in the high-\ac{SNR} regime, and when $K=0$. Unlike in the previous scenario, there is no range of SNR values on which correlation improves secrecy capacity. We also note that neglecting the impact of correlation on the average SNR at Eve would lead to the incorrect conclusion that correlation is beneficial for physical layer security in this scenario, as indicated by the curves in dashed lines.

\begin{figure}[t]
	\centering
		\includegraphics{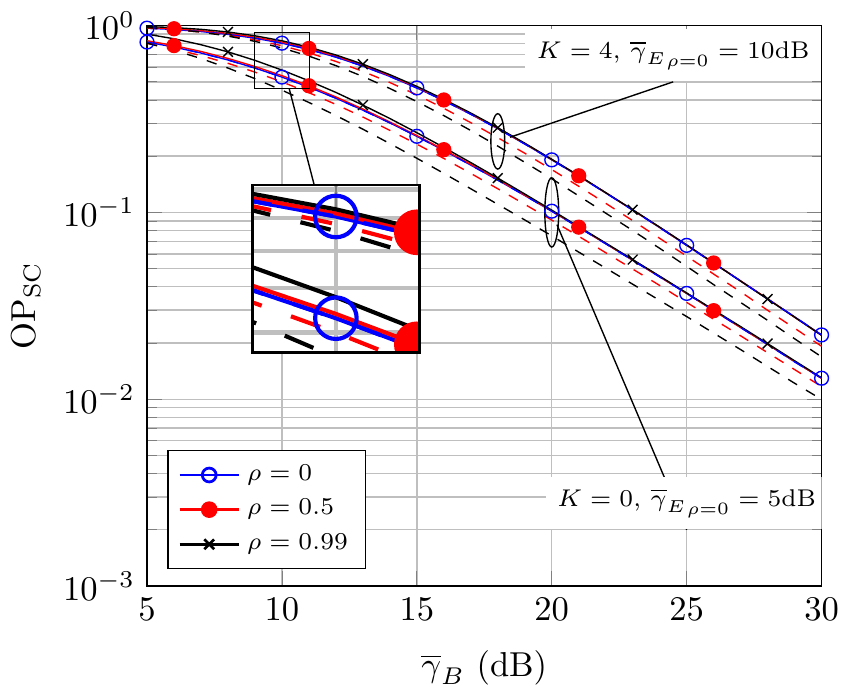}
	\caption{\ac{OP} of secrecy capacity $\overline{C}_{\rm S}$ vs. $\overline\gamma_B$, for different values of the correlation coefficient $\rho$ and \ac{LOS} conditions, with $R_{\rm th}=1$ bps/Hz. Solid lines correspond to the case on which $\overline\gamma_E=\{5,10\}$dB for $\rho=0$, and then enlarged by a factor $\Delta_{\rm PO}$. Dashed lines indicate the incorrect case where $\overline\gamma_E=\{5,10\}$ dB regardless of $\rho$. Markers correspond to \ac{MC} simulations, and the theoretical expressions in \eqref{eqopscB} and \eqref{eqopscB2} have been used to generate the dashed and solid line plots.} 
	\label{fig:8}
\end{figure}

Let us now move to the evaluation of the \ac{OP} of secrecy capacity in Fig. \ref{fig:8}. We use the same set of parameters as in Fig. \ref{fig:7}, and a secrecy outage rate threshold $R_{\rm th}=1$ bps/Hz. We now see that in the high-\ac{SNR} regime, i.e., in the low \ac{OP} operation zone, the secrecy performance is virtually agnostic to correlation. It is only when we move to higher values of \ac{OP}, i.e., lower values of the \ac{SNR} at Bob, when we can notice that correlation is slightly detrimental for physical layer security (see zoomed area in the figure). We also note that almost no effects are observed for the \ac{LOS} configuration, compared to the case of all-Rayleigh fading. Finally, we observe that the \ac{OP} of secrecy capacity that would be obtained by neglecting the effect of correlation on the average SNR at Eve would again induce the false intuition that correlation improves physical layer security performance -- which is not the case in this second scenario.

\section{Conclusions}
\label{Conclusions}
We investigated the effect of correlation between the energy and information links in \ac{EH} devices in the context of wireless powered communications. By rigorously quantifying the impact of correlation on the capacity of a WPC link, we showed that correlation causes an effective increase on the mean and variance of the receive SNR for a fixed transmit power -- or similarly, it allows to reduce the system power budget for a target average SNR at the receiver. In the latter case, correlation degrades capacity of a point-to-point WPC link. Now, when considering the effect of correlation on a physical layer security set-up, we showed that in some cases correlation turns out being beneficial from a secrecy capacity perspective. We also established that neglecting the impact of correlation on the average SNRs at the legitimate and eavesdropper's ends may cause that incorrect conclusions are extracted related to the role of correlation between the energy and information links for physical layer security. Future research activities in this line will include the consideration of non-ideal energy harvesting devices, as well as the derivation of reasonably tractable analytical expressions for the secrecy performance metrics, a rather intricate problem from a mathematical point of view.

\section*{Acknowledgements}
The authors gratefully acknowledge the help of Eduardo Martos Naya on the definition of the random variable transformations used in this paper.
\bibliographystyle{IEEEtran}
\bibliography{refs}

\end{document}